\documentclass[aps,prd,twocolumn,superscriptaddress,nofootinbib,preprintnumbers,longbibliography,floatfix]{revtex4-2}

\usepackage[utf8]{inputenc}
\usepackage[T1]{fontenc}
\usepackage{amsmath,amssymb,amsfonts}
\usepackage{graphicx}
\usepackage{xcolor}
\usepackage{microtype}
\usepackage{hyperref}

\hypersetup{
    colorlinks=true,
    linkcolor=blue,
    citecolor=blue,
    urlcolor=blue
}
\usepackage[capitalize]{cleveref}

\begin{document}


\title{Fisher-Rao Geometry of Gluon Saturation Fronts}

\author{Edoardo Spezzano}
\email{edoardo.spezzano@uni-muenster.de}
\affiliation{Department of Natural Sciences, Baruch College, City University of New York, New York, NY 10010, USA}

\date{\today}

\begin{abstract}
How nonlinear gluon saturation erases the memory of initial conditions is a central question in high-energy QCD. We develop the Fisher-Rao information geometry of saturation fronts in the Balitsky-Kovchegov (BK) equation and its Fisher-KPP diffusive limit. Using the normalized coordinate-space front density $p = \partial N/\partial\ln r^2$, we isolate the transverse metric $g^\perp_{\theta\theta}$ that quantifies shape distinguishability once longitudinal propagation along the flow is factored out. Nonlinear saturation arrests diffusive broadening and freezes the Fisher speed. Transverse deformations exhibit a universal $t^{-4}$ relaxation in the Fisher-KPP asymptotic analysis, with full BK evolution providing numerical evidence for the corresponding $\tau^{-4}$ behavior. We trace this steep power law to a relaxation mechanism: deforming the initial profile is equivalent, to leading order, to a spatial translation combined with an effective time delay of the universal relaxation flow. Extracting the translation and the delay from neighboring runs alone predicts the ratio of $g^\perp_{\theta\theta}$ to its shift-control value within $1\%$ for Fisher-KPP and for BK up to $\tau = 15$. For Fisher-KPP, the asymptotic front expansion yields the $t^{-4}$ law conditionally on stated analytic hypotheses, with a universal amplitude $C_*$ that agrees with numerical extrapolations to sub-percent accuracy. For BK, the asymptotic translation information exceeds its diffusive approximation by a factor of 3.5, which we attribute to the non-diffusive Levin-Tuchin saturation region. These results provide an information-geometric characterization of the approach to geometric scaling, consistent with the persistent sensitivity of pre-asymptotic HERA fits to initial shape parameters, and motivate systematic studies of parameter sensitivity in EIC kinematics.
\end{abstract}

\maketitle

\section{Introduction}

The emergence of high-energy gluon saturation in Quantum Chromodynamics (QCD) can be formulated as the propagation of nonlinear traveling wave fronts~\cite{Munier:2003gx, Munier:2003sj}. The Balitsky-Kovchegov (BK) equation~\cite{Balitsky:1995ub, Kovchegov:1999yj, Kovchegov:2000hz} belongs to the universality class of pulled fronts of the Fisher-Kolmogorov-Petrovsky-Piskunov (FKPP) reaction-diffusion equation~\cite{Fisher:1937, Kolmogorov:1937}, which provides the dynamical origin of geometric scaling~\cite{Stasto:2000er}. In nonlinear wave theory, the convergence of such fronts to a universal profile with systematically controlled corrections is well established~\cite{Bramson:1983, vanSaarloos:2000, Berestycki:2017, Graham:2017}.

Information-theoretic concepts have a fruitful history in small-$x$ physics. Peschanski introduced a dynamical entropy from the relative entropy between gluon distributions at different rapidities~\cite{Peschanski:2012cw}, whose small-rapidity limit is the Fisher information of a translating profile. This framework was extended to hadronic and nuclear collisions in Refs.~\cite{Ramos:2022, Ramos:2025}. More recently, Fisher-Rao geometry has been applied to linear DGLAP evolution in two recent preprints: Kou and Chen showed that nonsinglet distributions trace paths close to small circles on the Hilbert sphere~\cite{Kou:2026dglap}, while Bondarenko and Prygarin derived a closed-form lower bound on the Fisher speed from the mean drift and width of the logarithmic momentum fraction~\cite{Bondarenko:2026dglap}, a bound of the Cramér-Rao type discussed by Ito and Dechant~\cite{Ito:2020}.

These earlier investigations followed a single evolution trajectory. In the nonlinear BK regime, however, distinct initial conditions contract onto a common orbit of traveling fronts. A central geometric question is therefore how distinguishability between trajectories decays once the shared longitudinal motion along the flow is factored out. In this work, we separate the longitudinal Fisher speed $\sqrt{g_{\tau\tau}}$ from the transverse metric component $g^\perp_{\theta\theta}$, and introduce a rigid shift control that isolates the flow-induced relaxation.
Our main numerical finding, for both the FKPP benchmark and full coordinate-space BK evolution at fixed coupling, is that the transverse deformation is dominated by a single relaxation mechanism: to leading order, varying the initial condition is equivalent to a spatial translation of the front combined with a temporal delay of its universal relaxation. We demonstrate this mechanism both via the collinearity of the projected tangent residuals and by predicting the amplitude of $g^\perp_{\theta\theta}$ from profiles of neighboring runs alone. This mechanism accounts for the observed $g^\perp_{\theta\theta} \propto t^{-4}$ power-law decay. For FKPP, we derive this scaling and its universal amplitude $C_*$ from the asymptotic expansion of the front~\cite{Nolen:2019, Graham:2017} (\cref{app:t4}). The derivation is conditional: we state the analytic hypotheses it requires, which are not proven, and test its predictions numerically. For BK, the $\tau^{-4}$ behavior is a numerical observation, without a corresponding asymptotic derivation. We also evaluate the asymptotic translation information and the mean-drift bound for the BK front, showing that the front interior breaks the diffusive approximation by a factor of 3.5.

\section{Information-Geometric Framework}
\subsection{BK Equation at Fixed Coupling}
In coordinate space for a homogeneous target, the leading-order BK equation for the dipole amplitude $N(r,\tau)$ reads
\begin{multline}
\partial_\tau N(r) = \frac{1}{2\pi} \int d^2z \frac{r^2}{z^2(r-z)^2} \big[ N(z) + N(|r-z|) \\
- N(r) - N(z)N(|r-z|) \big],
\label{eq:BK}
\end{multline}
with $\tau \equiv \bar{\alpha}Y$ and $\bar{\alpha} = \alpha_s N_c/\pi$. The amplitude interpolates between color transparency ($N \to 0$ for $r \to 0$) and the black-disc limit ($N \to 1$ for $r \gg 1/Q_s$). We employ the Golec-Biernat-Wüsthoff (GBW) and McLerran-Venugopalan (MV) initial conditions~\cite{GolecBiernat:1998js, McLerran:1993ni, McLerran:1994ka} in their anomalous-dimension forms:
\begin{equation}
\begin{aligned}
N_{\rm GBW^\gamma} &= 1 - \exp\left[ -\frac{(r^2Q_{s0}^2)^\gamma}{4} \right], \\
N_{\rm MV^\gamma} &= 1 - \exp\left[ -\frac{(r^2Q_{s0}^2)^\gamma}{4} \ln\left(\frac{1}{r\Lambda} + e\right) \right],
\end{aligned}
\label{eq:ic}
\end{equation}
with $\Lambda = 0.241\,Q_{s0}$; $\gamma = 1$ recovers the standard models. Solutions approach geometric scaling,
\begin{equation}
N(r,\tau) \simeq \mathcal{N}\big(r^2 Q_s^2(\tau)\big),
\label{eq:scaling}
\end{equation}
with
\begin{equation}
\ln Q_s^2 = v_c \tau - \frac{3}{2\gamma_c} \ln \tau + \dots
\label{eq:Qs}  
\end{equation}
Here $\chi(\gamma) = 2\psi(1) - \psi(\gamma) - \psi(1-\gamma)$ is the leading-order BFKL eigenvalue~\cite{Kuraev:1977fs, Balitsky:1978}, $\gamma_c \simeq 0.6275$ solves $\chi'(\gamma_c) = \chi(\gamma_c)/\gamma_c$, and $v_c = \chi(\gamma_c)/\gamma_c \simeq 4.883$~\cite{Mueller:2002zm, Munier:2003sj}. We define the saturation scale $Q_s(\tau)$ by $N(1/Q_s, \tau) = 1/2$; it differs from the input parameter $Q_{s0}$ of \cref{eq:ic}.

\subsection{The Front Density}
Using the logarithmic coordinate $\rho \equiv \ln (r^2 Q_{\rm ref}^2)$ with a fixed reference scale $Q_{\rm ref}$, equal to $Q_{s0}$ of the unperturbed initial condition, we define the front density
\begin{equation}
p(\rho \mid \tau, \theta) \equiv \frac{\partial N(\rho, \tau; \theta)}{\partial \rho}, \quad
\int_{-\infty}^{+\infty} p \, d\rho = 1.
\label{eq:density}
\end{equation}
Because $N$ is monotonic, $p \ge 0$ is a normalized probability density on $\mathbb{R}$, localizing where the saturation transition occurs in dipole size without infrared or ultraviolet cutoffs. Unlike momentum space, where the black-disc limit produces an unnormalizable logarithmic divergence at small $k$, the coordinate-space density $p$ is strictly normalizable. While $p$ is not the distribution of an experimental observable, the resulting Fisher-Rao metric provides an intrinsic Riemannian measure of distinguishability between front profiles.

The front sits at $\rho = -m(\tau)$, where
\begin{equation}
m(\tau) \equiv \ln\frac{Q_s^2(\tau)}{Q_{\rm ref}^2}, \qquad \sigma \equiv -m,
\label{eq:m}
\end{equation}
so that $\sigma$ is the signed front position in $\rho$ and $\dot m = d\ln Q_s^2/d\tau = -\dot\sigma$ is the front velocity.

\subsection{Fisher-Rao Metric and Key Identities}
On the statistical manifold parameterized by coordinates $\xi^i = (\tau, \theta)$, where $\tau$ is the evolution variable and $\theta$ parameterizes the initial condition, the Fisher-Rao metric~\cite{Amari:2000} is
\begin{equation}
g_{ij} = \int_{-\infty}^{+\infty} \frac{1}{p} \frac{\partial p}{\partial \xi^i} \frac{\partial p}{\partial \xi^j} d\rho.
\label{eq:Fisher_metric}
\end{equation}
Three standard identities govern the analysis:

\noindent\textit{Identity 1 (Translating families):}
For a traveling profile
$p(\rho \mid \tau) = f(\rho \pm m(\tau))$,
the longitudinal metric component reduces to
\begin{equation}
g_{\tau\tau} = \dot{m}^2 I[f], \qquad
I[f] \equiv
\int_{-\infty}^{+\infty}
\frac{[f'(\xi)]^2}{f(\xi)}\,d\xi,
\label{eq:translating}
\end{equation}
where $\xi = \rho \pm m(\tau)$ is the comoving
coordinate and $I[f]$ is the Fisher translation
information of the profile.

\noindent\textit{Identity 2 (Schur complement decomposition):}\\
For $g_{\tau\tau} > 0$, the transverse metric component is defined by the Schur complement
\begin{equation}
g_{\theta\theta}^\perp \equiv g_{\theta\theta} - \frac{g_{\tau\theta}^2}{g_{\tau\tau}} = \min_c \left\| \partial_\theta p - c\, \partial_\tau p \right\|_F^2,
\label{eq:orthogonal_metric}
\end{equation}
where $\|h\|_F^2 = \int h^2/p \, d\rho$ and $c = g_{\tau\theta}/g_{\tau\tau}$. Here $c\,\partial_\tau p$ is the optimal longitudinal projection (the shift in evolution time that best mimics a change in $\theta$), and $g^\perp_{\theta\theta}$ measures the genuine shape residual. A decaying $g^\perp_{\theta\theta}$ signals that trajectories become indistinguishable up to a time translation.

\noindent\textit{Identity 3 (Mean-drift bound):}
Applying the Cauchy-Schwarz inequality to $\rho$ yields the bound~\cite{Ito:2020, Bondarenko:2026dglap}
\begin{equation}
g_{\tau\tau} \ge \frac{(d\langle\rho\rangle/d\tau)^2}{\sigma_\rho^2}, \qquad \mathcal{R}_B \equiv \frac{g_{\tau\tau}\,\sigma_\rho^2}{(d\langle\rho\rangle/d\tau)^2} \ge 1,
\label{eq:meanbound}
\end{equation}
where $\langle\rho\rangle$ and $\sigma_\rho^2$ are the mean and variance of $\rho$ under $p$. For a pure translating front, $\mathcal{R}_B = I[f]\,\mathrm{Var}[f]$ is the Cramér-Rao product, which equals 1 only for a Gaussian.

\subsection{Shape Parameters and Shift Control}
\label{sec:control}
We investigate two classes of parameters $\theta$: shape parameters ($\gamma_0$ for FKPP; $\gamma$ for BK) that alter the profile form, and a shift parameter $\theta_s$ that translates the initial condition rigidly: $u_0(x - \theta_s)$ for FKPP and $N_0(\rho + \theta_s)$ for BK. For BK this is equivalent to scaling $Q_{s0}^2 \to Q_{s0}^2 e^{\theta_s}$ in the initial condition only, with $Q_{\rm ref}$ held fixed, so that $\theta_s$ translates the distribution in $\rho$ rather than redefining the coordinate. Translation invariance implies
\begin{equation}
\partial_{\theta_s} p = +\partial_\rho p \ \ (\text{BK}), \qquad \partial_{\theta_s} p = -\partial_x p \ \ (\text{FKPP}).
\label{eq:shift}
\end{equation}
For a front that propagates as a rigidly translating profile, $p = f(\rho + m(\tau))$, one has $\partial_{\theta_s} p \propto \partial_\tau p$ and hence $g^\perp_{\rm shift} = 0$. At finite times the profile is still relaxing, so $g^\perp_{\rm shift}$ is generally nonzero: it isolates the component of $g^\perp_{\theta\theta}$ induced solely by the non-steadiness of the flow, and its decay is the subject of our analysis.

\section{Numerical Methods}
\label{sec:numerics}

Both models are evolved together with their forward tangent equations, so that $\partial_\theta p$ is obtained at the accuracy of $p$ itself, while $\partial_\tau p$ follows from the evolution operator. The transverse component is evaluated directly as the Fisher norm of the projected residual $\partial_\theta p - c\,\partial_\tau p$, which avoids the cancellation in the Schur complement, and the Fisher integrands are truncated in the far tails, where $p$ is dominated by round-off. The front position is the level crossing $N = 1/2$ ($u = 1/2$ for FKPP), and every velocity is the exact time derivative of that crossing. FKPP [\cref{eq:FKPP}] is integrated with fourth-order finite differences and RK4 up to $t = 1280$.

For BK, the kernel symmetry $z \leftrightarrow r-z$, implemented through the partition of unity $w(z) + w(r-z) = 1$ with $w(z) = |r-z|^2/(z^2 + |r-z|^2)$, yields
\begin{equation}
\partial_\tau N(r) = \frac{1}{\pi} \int du \int_0^\pi d\phi\, \frac{B(r, z)}{1 + 2x^2 - 2x\cos\phi},
\label{eq:BK_disc}
\end{equation}
where $B$ is the BK bracket in \cref{eq:BK}, $u = \ln(z^2/r^2)$, and $x = e^{u/2}$. On a uniform $\rho$-grid this operator is translation invariant, so the evolution reduces to discrete convolutions and a bilinear form in shifted copies of $N$. We use $\rho \in [-420, 10]$ with $\Delta\rho = 0.15$, wide enough to contain the diffusive leading edge up to $\tau = 40$, $u \in [-40, 30]$, and RK4 with $d\tau = 0.02$.

The longitudinal observables $\dot m$, $I_\rho$, and $\mathcal{R}_B$ are converged to $10^{-5}$. The transverse component, which reaches $10^{-9}$ for BK and $10^{-15}$ for FKPP, is converged to $0.3\%$ for FKPP once saturated-side round-off is removed, and to about $2\%$ for BK at $\tau = 40$, where the time step and the upper limit of $u$ dominate the uncertainty. 

\section{Fisher-KPP Dynamics: Freezing and Memory Loss}
\label{sec:fkpp}

In the diffusion approximation, the BK equation reduces to the FKPP equation~\cite{Munier:2003gx, Munier:2003sj}
\begin{equation}
\partial_t u = \partial_x^2 u + u - u^2, \quad p = -\partial_x u.
\label{eq:FKPP}
\end{equation}
We examine initial data $u_0 = \min(1, e^{-\gamma_0 x})$ with $\gamma_0 > 1$, for which fronts are pulled and their position advances as~\cite{Bramson:1983, vanSaarloos:2000, Nolen:2017, Nolen:2019, Graham:2017, Berestycki:2017}
\begin{equation}
\sigma(t) = 2t - \frac{3}{2}\ln t + \alpha_0 - \frac{3\sqrt{\pi}}{\sqrt{t}} + \mu_* \frac{\ln t}{t} + \frac{\alpha_1}{t} + \dots,
\label{eq:sigma}
\end{equation}
with $\mu_* = 9(5 - 6\ln 2)/8$. The non-universal constants $\alpha_0(\gamma_0)$ and $\alpha_1(\gamma_0)$ encode the memory of the initial condition; all other terms are universal. We take $\sigma(t)$ to be the position of the level $u = 1/2$; another level changes $\alpha_0$ and $\alpha_1$ only by universal constants, which drop out of their $\theta$-derivatives.

\subsection{Linear vs.\ Nonlinear Regime}
In linear diffusion with growth, the density $p(x,t)$ is a spreading Gaussian, so the Fisher speed decays as $t^{-1}$, driven purely by diffusive broadening. Nonlinear saturation arrests this spreading: $g_{tt}$ freezes at $4 I_x$, where $I_x \equiv I[f] = 0.087185$ is the translation information [\cref{eq:translating}] of the density $f = -\phi'$ of the critical traveling wave. The Cramér-Rao product $I_x\,\mathrm{Var}[f] = 1.1502$ exceeds unity, confirming that the asymptotic front is non-Gaussian.

\begin{figure*}[tbp]
\centering
\includegraphics[width=0.72\textwidth]{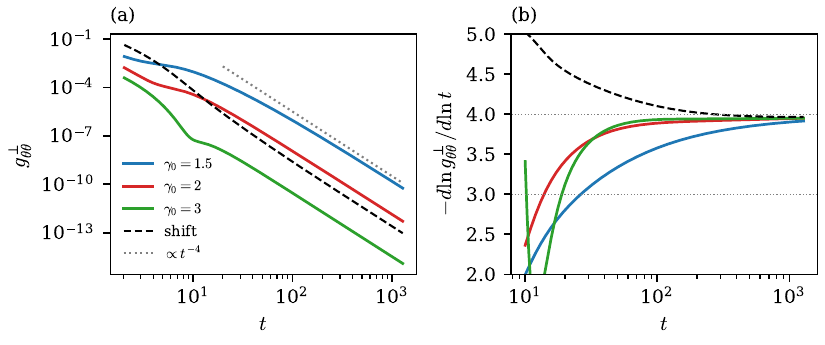}
\caption{FKPP dynamics ($\Delta x = 0.05$). (a) Transverse component $g^\perp_{\theta\theta}$ for $\theta = \gamma_0$ (solid) and shift control (dashed) alongside a $t^{-4}$ guide. (b) Local exponent $\beta_{\text{loc}} = -d\ln g^\perp_{\theta\theta}/d\ln t$, converging towards 4.}
\label{fig:fkpp}
\end{figure*}

\subsection{Transverse Component and Power-Law Decay}
\Cref{fig:fkpp} shows $g^\perp_{\theta\theta}$ for $\theta = \gamma_0$. Over five orders of magnitude of decay, the full metric $g_{\theta\theta} \to (\partial_\theta \alpha_0)^2 I_x$ approaches a finite constant while the transverse part vanishes. The local exponent $\beta_{\text{loc}} = -d\ln g^\perp_{\theta\theta}/d\ln t$ increases steadily towards 4 for all initial conditions, reaching $3.89$--$3.95$ at $t = 1280$, while that of the shift control stays within $0.08$ of 4 throughout.

This power law is accounted for by a formal asymptotic derivation from the front expansion~\cite{Nolen:2019, Graham:2017}, detailed in \cref{app:t4}. In the comoving frame $\xi = x - \sigma(t)$, the field expands as $u = \phi(\xi) + \psi(\xi)/t + \dots$, where $\phi$ and $\psi$ are universal profiles. The initial condition enters only through the position constants $\alpha_0$ and $\alpha_1$ up to order $t^{-2}$, with $\alpha_1$ acting as an effective time shift of the universal relaxation $\psi/t$. Writing $\partial_t p = A - \dot\sigma\,\partial_x p$ with $A \equiv \partial_t p|_\xi$ the comoving relaxation rate, one finds $t^2 A \to -Q$ where $Q = -\psi'$, yielding the scaling law
\begin{equation}
g^\perp_{\rm shift} \simeq \frac{C_*}{4t^4}, \qquad g^\perp_{\theta\theta} \simeq a_\infty'^2\,\frac{C_*}{4t^4}.
\label{eq:t4law}
\end{equation}
Here $C_*$ is the universal constant
\begin{equation}
C_* \equiv \left\| (1 - \Pi_{\widehat{D}_0}) \widehat{Q} \right\|_{L^2}^2 = \int \frac{\left[(1-\Pi_{f'})Q\right]^2}{f} \, d\xi \simeq 1.112,
\label{eq:Cstar}
\end{equation}
where $f = -\phi'$, $\widehat{Q} \equiv Q/\sqrt{f}$, and $\widehat{D}_0 \equiv f'/\sqrt{f}$ is the asymptotic shift tangent in the $L^2$ spherical embedding (with $\Pi_{f'}$ the corresponding Fisher projection). The amplitude coefficient is $a'_\infty = \partial_\theta\alpha_0 + \tfrac43\partial_\theta\alpha_1$, and the effective time delay is $t_0' = -\tfrac23\partial_\theta\alpha_1$. The derivation is asymptotic and conditional on the hypotheses listed in \cref{app:t4}; its predictions are tested numerically below.

At $t = 1280$, $t^4 g^\perp_{\rm shift}$ lies $10\%$ below the asymptotic limit $C_*/4$, consistent with the slow $O(t^{-1/2})$ approach governed by the $t^{-3/2}$ correction in the front position. As detailed in \cref{app:t4}, fitting this approach in powers of $t^{-1/2}$ yields an extrapolated coefficient in sub-percent agreement with $C_*/4$, and the delay relation $\partial_\theta\alpha_1 = -\tfrac32 t_0'$ is confirmed to within $1.2\%$.

\begin{figure*}[tbp]
\centering
\includegraphics[width=0.88\textwidth]{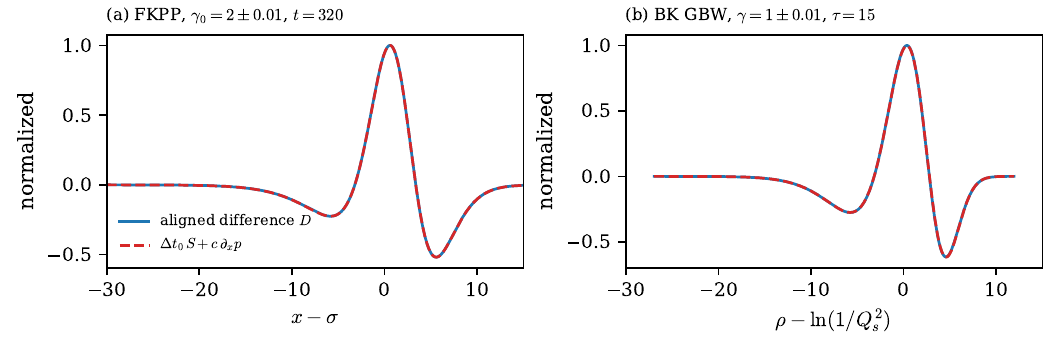}
\caption{Density difference $D$ between aligned runs at $\theta \pm 0.01$ (solid) and two-parameter fit $\Delta t_0\, S + c\,\partial_x p$ (dashed), normalized to $\max|D|$, with $S \equiv \partial_t p|_\xi$ the comoving shape relaxation rate. (a) FKPP, $\gamma_0 = 2$, $t = 320$ (residual $0.04\%$). (b) BK with GBW initial condition, $\gamma = 1$, $\tau = 15$ (residual $1\%$).}
\label{fig:delay}
\end{figure*}

\begin{table}[tbp]
\centering
\caption{Independent test of the delay relation, \cref{eq:delay}. Here $a' = \partial_\theta\sigma - \dot\sigma\, t_0'$ is extracted only from the profiles of runs at $\theta \pm 0.01$ (for BK, $\theta = \gamma$); Pred.\ is $a'^2$, Meas.\ is $g^\perp_{\theta\theta}/g^\perp_{\rm shift}$ from the tangent equation, and Dev.\ is Pred./Meas.${}-1$. }
\label{tab:delay}
\begin{ruledtabular}
\footnotesize
\begin{tabular}{lcccc}
Case & $a'$ & Pred. & Meas. & Dev. \\
\hline
\multicolumn{5}{l}{FKPP ($\gamma_0$, $t$)} \\
$1.5$, $320$ & $+23.48$ & $551.2$ & $550.2$ & $+0.17\%$ \\
$2.0$, $320$ & $+2.291$ & $5.248$ & $5.245$ & $+0.06\%$ \\
$3.0$, $320$ & $+0.1131$ & $0.01278$ & $0.01278$ & $+0.05\%$ \\
$1.5$, $1280$ & $+24.98$ & $624.0$ & $621.6$ & $+0.38\%$ \\
$2.0$, $1280$ & $+2.350$ & $5.520$ & $5.558$ & $-0.67\%$ \\
\hline
\multicolumn{5}{l}{BK ($\tau$)} \\
GBW, $15$ & $-1.323$ & $1.751$ & $1.738$ & $+0.74\%$ \\
MV, $15$ & $-2.744$ & $7.531$ & $7.476$ & $+0.74\%$ \\
GBW, $40$ & $-1.308$ & $1.710$ & $1.765$ & $-3.1\%$ \\
MV, $40$ & $-2.729$ & $7.450$ & $7.668$ & $-2.8\%$ \\
\end{tabular}
\end{ruledtabular}
\end{table}

\subsection{Direction of the Residual and the Universal Delay Mechanism}
\label{sec:delay}

The projected tangent residuals associated with variations of the initial shape are nearly collinear with those generated by the rigid-shift control. This numerical observation indicates that, to leading order, changing the initial condition produces a combination of spatial displacement and temporal delay of the universal relaxation dynamics.

In terms of the evolved density, this behavior is described by
\begin{equation}
\partial_\theta p \simeq t_0'\,\partial_t p - a'\,\partial_q p,
\label{eq:delaymodel}
\end{equation}
where the spatial derivative is understood as $\partial_x$ for FKPP and $\partial_\rho$ for BK, with $a' = \partial_\theta\sigma - \dot\sigma\, t_0'$ and $t_0' \equiv \partial_\theta t_0$. After removing the longitudinal component along $\partial_t p$, and noting that $-\partial_q p$ generates the rigid-shift tangent [\cref{eq:shift}], this leads directly to
\begin{equation}
g_{\theta\theta}^\perp \simeq a'^2\, g^\perp_{\rm shift}.
\label{eq:delay}
\end{equation}

We test this relation independently by extracting the displacement and delay from neighboring simulations at $\theta \pm 0.01$, without using the tangent equation to construct the prediction. After aligning the front positions, the difference $D$ between neighboring profiles is fitted in the Fisher norm by $\Delta t_0\, S + c\,\partial_q p$, where $S \equiv \partial_t p|_\xi$ is the comoving shape relaxation rate (\cref{fig:delay}). The front velocity $\dot\sigma$ is evaluated at the same level-$1/2$ crossing used for the position. As shown in \cref{tab:delay}, the resulting estimate agrees with the measured transverse metric at the percent level for FKPP and for BK at intermediate rapidities ($\tau = 15$). At larger BK rapidities ($\tau = 40$), the remaining discrepancies are consistent with the numerical sensitivity of the small transverse component.

\section{Full BK Evolution}

\subsection{Front Velocity and Longitudinal Diagnostics}
\begin{figure*}[tbp]
\centering
\includegraphics[width=0.88\textwidth]{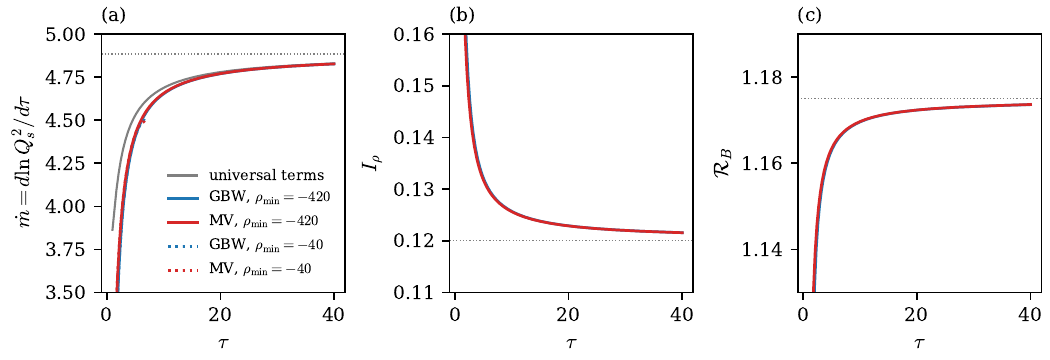}
\caption{BK evolution at fixed coupling. (a) Front velocity $\dot m = d\ln Q_s^2/d\tau$, compared with Ebert-van Saarloos asymptotics. (b) Translation information $I_\rho = g_{\tau\tau}/\dot m^2$ approaching $0.120$. (c) Mean-drift bound ratio $\mathcal{R}_B$ approaching $1.175$.}
\label{fig:bk_front}
\end{figure*}

The front velocity $\dot m = d\ln Q_s^2/d\tau$ converges towards $v_c \simeq 4.883$ from below, reaching $4.731$ at $\tau = 15$ and $4.826$ at $\tau = 40$, within $0.06\%$ of the universal Ebert-van Saarloos asymptotics~\cite{vanSaarloos:2000} at $\tau = 40$ [\cref{fig:bk_front}(a)]. Because the BK front is pulled, its velocity is selected in the leading edge. Truncating the grid at $\rho_{\min} = -40$ changes $\dot m$ by only $1$--$2\%$ at $\tau = 7$ but $g^\perp_{\theta\theta}$ by $30$--$40\%$: transverse distinguishability is far more sensitive to the leading edge than the front velocity.

\subsection{Translation Information and the Diffusive Breakdown}
The translation information $I_\rho \equiv g_{\tau\tau}/\dot{m}^2$ converges towards the asymptotic value $I_\rho^{\rm BK} \simeq 0.120$, identically for both GBW and MV initial conditions [\cref{fig:bk_front}(b)].

In the diffusive approximation, expanding $\chi(\gamma)$ around $\gamma_c$ maps BK onto FKPP via $x = \gamma_c \rho$, predicting $I_\rho^{\rm diff} = \gamma_c^2 I_x = 0.0343$. This underestimates the full BK value by a factor of $3.5$, which we attribute to the different front interiors of FKPP and BK. Behind the BK front, deep in the saturation region $\ln(r^2Q_s^2) \gg 1$, the BK equation for the scattering matrix $S = 1 - N$ reduces to $\partial_\tau \ln S \simeq -\ln(r^2Q_s^2)$~\cite{Levin:2000}. With $\ln(r^2Q_s^2) = \rho + m(\tau)$ and $\dot m \to v_c$, this gives the Levin-Tuchin law~\cite{Levin:2000}
\begin{equation}
1 - N \simeq \exp\left[-\frac{\ln^2(r^2Q_s^2)}{2v_c}\right] = \exp\left[-\frac{\gamma_c\,(\rho + m)^2}{2\chi(\gamma_c)}\right].
\label{eq:LT}
\end{equation}
This describes a Gaussian falloff of the $S$-matrix with a width independent of rapidity. Our profiles at $\tau = 40$ follow it: for $3 < \ln(r^2Q_s^2) < 16$, a quadratic fit of $\ln(1-N)$ has curvature $0.106$ for both GBW and MV, compared with $1/(2v_c) = 0.102$. The critical FKPP wave instead saturates exponentially, $1 - \phi \propto e^{(\sqrt2-1)\xi}$ as $\xi \to -\infty$. Since the Fisher translation information $I[f]$ integrates over the entire front profile, this structural difference in the saturation interior accounts for the failure of the diffusive approximation to reproduce the BK value.

\subsection{Mean-Drift Bound}
The ratio $\mathcal{R}_B = g_{\tau\tau}\sigma_\rho^2 / (d\langle\rho\rangle/d\tau)^2 \ge 1$ diagnoses non-Gaussianity and non-steady deformation. Being scale-invariant under $\rho \to \lambda \rho$, $\mathcal{R}_B$ enables a direct parameter-free comparison between models. For BK, $\mathcal{R}_B$ approaches $\mathcal{R}_B^{\rm BK} \simeq 1.175$ identically for GBW and MV [\cref{fig:bk_front}(c)], consistent with a universal front. For critical FKPP fronts $\mathcal{R}_B^{\rm FKPP} \simeq 1.1502$, so the excess over the Gaussian bound is $17\%$ larger in BK, reflecting the non-diffusive front interior discussed above.

\subsection{Transverse Component in BK}
\label{sec:bk_perp}
\begin{figure*}[tbp]
\centering
\includegraphics[width=0.72\textwidth]{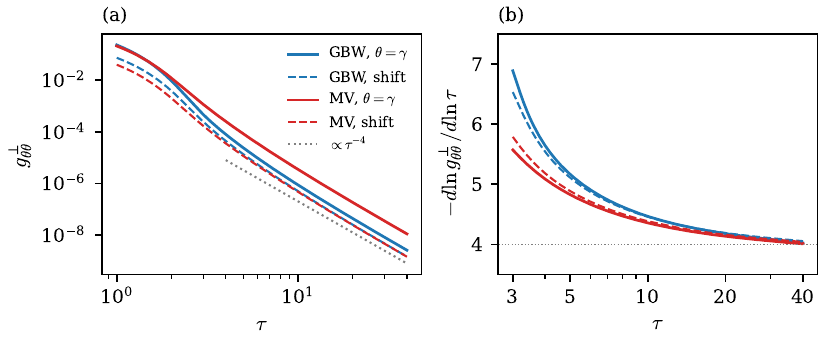}
\caption{BK transverse component. (a) $g^\perp_{\theta\theta}$ for $\theta = \gamma$ (solid) and shift control (dashed) alongside a $\tau^{-4}$ guide. (b) Local exponent $-d\ln g^\perp_{\theta\theta}/d\ln \tau$, approaching 4.}
\label{fig:bk_memory}
\end{figure*}

As shown in \cref{fig:bk_memory}, $g^\perp_{\theta\theta}$ decays by approximately eight orders of magnitude over the evolution, tracking the $\tau^{-4}$ guide at late times. The local exponent $-d\ln g^\perp_{\theta\theta}/d\ln\tau$ decreases towards 4, reaching $4.02$ for GBW (and $4.01$ for MV) at $\tau = 40$, and shifting to $4.06$ upon halving the time step. Meanwhile, the rescaled shift control $\tau^4 g^\perp_{\rm shift}$ converges towards a common value for both GBW and MV initial conditions [\cref{fig:bk_memory}(a)]. The Fisher cosine between the shape-parameter residual and the shift residual is $0.99993$ at $\tau = 15$ and approaches $0.9999$ under numerical refinement at $\tau = 40$, indicating that the delay mechanism operates in full BK evolution as in FKPP. These results provide strong numerical evidence for an approach to $\tau^{-4}$ relaxation in BK, though unlike FKPP, no asymptotic derivation supports them.

The independent test (\cref{tab:delay}) confirms $g^\perp_{\theta\theta} \simeq a'^2 g^\perp_{\rm shift}$ in BK within $0.75\%$ at $\tau = 15$; the $3\%$ deviation at $\tau = 40$ is numerical and drops to $0.55\%$ on a finer grid.
\newpage
\section{Discussion and Conclusions}

\subsection{Relation to DGLAP Evolution}
Three fundamental contrasts distinguish nonlinear saturation from linear DGLAP evolution:
(i)~\textit{Speed freezing}: in the linear DGLAP trajectories studied in Ref.~\cite{Bondarenko:2026dglap}, the Fisher speed decreases with scale within fixed-flavor segments and need not be monotonic across heavy-quark thresholds; nonlinear saturation instead arrests broadening and produces an asymptotically finite Fisher speed, $\dot m \sqrt{I[f]}$.
(ii)~\textit{Mean-drift bound}: for three of the five model inputs of Ref.~\cite{Bondarenko:2026dglap}, the mean-drift bound accounts for at least $94.8\%$ of the squared Fisher speed ($\mathcal{R}_B \le 1.055$), whereas for the asymptotic BK front it accounts for $85\%$ ($\mathcal{R}_B \simeq 1.175$), a departure from Gaussianity that we attribute to the Levin-Tuchin interior.
(iii)~\textit{Contraction of families}: DGLAP evolution is studied along single trajectories, whereas our numerical results support an effective contraction of families of initial profiles onto the universal traveling-front orbit, with the transverse metric $g^\perp_{\theta\theta}$ quantifying the loss of memory along the sampled directions.

\subsection{Phenomenological Considerations for Small-\texorpdfstring{$x$}{x} Data}
\label{sec:pheno}
The density $p(\rho)$ is an intrinsic geometric property of the dipole profile, not an experimental count rate. Experimental sensitivity is governed instead by the Fisher information matrix of the deep inelastic scattering (DIS) data likelihood,
\begin{equation}
F_{ab}(Y) = \sum_i \frac{1}{\delta_i^2} \frac{\partial \sigma_r(x_i, Q_i^2)}{\partial \theta_a} \frac{\partial \sigma_r(x_i, Q_i^2)}{\partial \theta_b},
\label{eq:data_fisher}
\end{equation}
where $\sigma_r$ is the reduced cross section and $\delta_i$ are independent experimental uncertainties (generalized to $F = J^T C^{-1} J$ in the presence of correlated systematic errors). Shape deformations suppressed in $p$ may also be suppressed in $F$, but this has to be checked by computing $F$, and the two need not decay at the same rate: DIS cross sections fold the dipole amplitude with the virtual photon wave function, sampling a localized window in dipole size $r \sim 1/Q$ determined by kinematics, rather than the density $p$ that defines our metric.

HERA data span $10^{-6} \lesssim x \lesssim 10^{-2}$, about nine units of rapidity, corresponding to $\tau \lesssim 2$ for $\bar\alpha \approx 0.2$. In this pre-asymptotic range $g^\perp_{\theta\theta}$ is still orders of magnitude larger than at $\tau \gtrsim 10$ (\cref{fig:bk_memory}), which is consistent with the persistent sensitivity of phenomenological fits with running coupling to initial shape parameters such as $\gamma$~\cite{Albacete:2010sy, Lappi:2013zma}. A quantitative connection requires the likelihood Fisher matrix of \cref{eq:data_fisher}, which we have not computed. At the Electron-Ion Collider (EIC), where the lever arm for saturation comes from nuclear enhancement ($Q_s^2 \propto A^{1/3}$) rather than large rapidity spans, whether this contraction reduces parameter sensitivity in observables with running coupling remains to be studied.

\subsection{Summary}
We have developed the Fisher-Rao information geometry of gluon saturation fronts in FKPP and BK evolution, decomposing the metric into a longitudinal flow speed and a transverse memory component $g^\perp_{\theta\theta}$. The principal conclusions are:

(i)~\textit{Speed freezing and non-diffusive invariants} (numerical observation, both models): Nonlinear saturation arrests diffusive broadening, freezing the longitudinal Fisher speed. For BK, the asymptotic translation information $I_\rho^{\rm BK} \simeq 0.120$ exceeds its diffusive counterpart $I_\rho^{\rm diff} \simeq 0.0343$ by a factor of $3.5$, and the mean-drift ratio $\mathcal{R}_B^{\rm BK} \simeq 1.175$ exceeds its FKPP counterpart $\mathcal{R}_B^{\rm FKPP} \simeq 1.1502$, reflecting the non-Gaussian Levin-Tuchin interior.

(ii)~\textit{Universal delay mechanism} (numerical validation, both models): Transverse deformations are dominated by a single relaxation mode: deforming the initial condition is equivalent, to leading order, to a spatial translation combined with a temporal delay of the universal relaxation dynamics. This yields $g^\perp_{\theta\theta} \simeq a'^2 g^\perp_{\rm shift}$, which reproduces the measured transverse metric from independent simulation profiles without tangent integration.

(iii)~\textit{Transverse scaling} (conditional derivation for FKPP; numerical evidence for BK): In FKPP, the delay mechanism leads to the asymptotic power-law decay $g^\perp_{\rm shift} \sim C_*/(4t^4)$ with universal amplitude $C_* \simeq 1.112$, derived conditionally from the front expansion and confirmed numerically. In full coordinate-space BK, numerical evolution reveals the corresponding $\tau^{-4}$ relaxation law, with the local decay exponent converging towards 4.

Open problems are a proof of these hypotheses; the corresponding asymptotic expansion for the full BK front, where the rescaled shift control already approaches a value common to both initial conditions; and the experimental likelihood Fisher matrix $F_{ab}$ under running coupling, to examine whether parameter sensitivity is reduced in deep inelastic scattering at the future Electron-Ion Collider.

\section*{Data Availability}
The data that support the findings of this article are not publicly available. The simulation code, the stored run outputs, and the scripts that generate every table and figure are available from the author upon reasonable request.

\begin{acknowledgments}
Anthropic's Claude Opus 5.5 was used to write simulation and analysis code under the author's direction. The code was validated against analytical results (the BFKL eigenvalue $\chi(\gamma)$, linear Gaussian diffusion, and the critical FKPP wave), against an independent adaptive-quadrature evaluation of the BK right-hand side, and by the convergence tests reported here. The author reviewed all AI-assisted material and takes full responsibility for the content.
\end{acknowledgments}

\appendix
\crefalias{section}{appendix}

\section{Asymptotic origin of the \texorpdfstring{$t^{-4}$}{t\textasciicircum -4} law in FKPP}
\label{app:t4}

\begin{figure*}[tbp]
\centering
\includegraphics[width=0.88\textwidth]{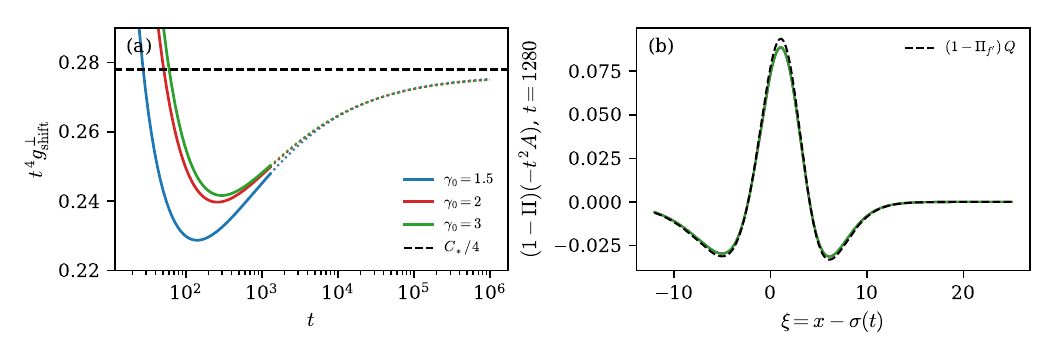}
\caption{Asymptotic predictions for FKPP ($\Delta x = 0.05$). (a) $t^4 g^\perp_{\rm shift}$ for three initial conditions (solid), fits $c_0 + c_1 t^{-1/2} + c_2 t^{-1} + c_3 t^{-3/2}$ extended beyond $t = 1280$ (dotted), and the predicted limit $C_*/4$ (dashed). (b) Transverse part of the rescaled relaxation $-t^2 A$ at $t = 1280$ for $\gamma_0 = 1.5, 2, 3$ (colors as in (a)) and the universal profile $(1-\Pi_{f'})Q$ from \cref{eq:psi} (dashed).}
\label{fig:graham}
\end{figure*}

\subsection{Exact reduction}
Under the spherical embedding $p \mapsto 2\sqrt{p}$~\cite{Amari:2000}, the Fisher norm of any tangent perturbation $h = \partial p$ is the standard $L^2$ norm of the embedded vector
\begin{equation}
\widehat{h} \equiv \frac{h}{\sqrt{p}} \in L^2(\mathbb{R}),
\label{eq:embedding}
\end{equation}
such that $\langle h_1, h_2 \rangle_F = \langle \widehat{h}_1, \widehat{h}_2 \rangle_{L^2} = \int \widehat{h}_1 \widehat{h}_2 \, d\rho$. Let $D = \partial_x p$ be the spatial shift tangent and $A = \partial_t p|_{\xi}$ the shape relaxation rate in the comoving frame $\xi = x - \sigma(t)$, satisfying
\begin{equation}
\partial_t p = A - \dot\sigma D.
\label{eq:A}
\end{equation}
In the embedded representation $\widehat{h} = h/\sqrt{p}$, dividing by $\sqrt{p}$ yields $\widehat{\partial_t p} = \widehat{A} - \dot\sigma \widehat{D}$. For a shape parameter, writing the tangent in delay form gives
\begin{equation}
\partial_\theta p = t_0'\,\partial_t p - a'\,D + \varepsilon \implies \widehat{\partial_\theta p} = t_0'\,\widehat{\partial_t p} - a'\,\widehat{D} + \widehat{\varepsilon},
\label{eq:delayform}
\end{equation}
where $\widehat{\varepsilon} \equiv \varepsilon/\sqrt{p}$ is the embedded remainder. Let $\Pi_t$ denote the orthogonal projection in $L^2(\mathbb{R})$ onto $\widehat{\partial_t p}$. Because $(1-\Pi_t)\widehat{\partial_t p} = 0$, the following algebraic identities hold exactly:
\begin{equation}
\begin{aligned}
(1-\Pi_t)\,\widehat{D} &= \dot\sigma^{-1}\,(1-\Pi_t)\,\widehat{A}, \\
(1-\Pi_t)\,\widehat{\partial_\theta p} &= -a'\,(1-\Pi_t)\,\widehat{D} + (1-\Pi_t)\,\widehat{\varepsilon} .
\end{aligned}
\label{eq:exact}
\end{equation}

To obtain the asymptotic scaling of the transverse metric, multiply \cref{eq:exact} by $t^2$. In the large-time limit, the front velocity approaches $\dot\sigma \to s_0$ (with $s_0 = 2$ for FKPP), and $\widehat{\partial_t p} \to -s_0 \widehat{D}_0 \neq 0$, so that the projection operator $\Pi_t$ converges strongly in $L^2$ to $\Pi_{\widehat{D}_0}$, the projection along the asymptotic shift tangent. If the comoving relaxation scales as $t^2 \widehat{A} \to \widehat{A}_0$, the shift tangent approaches $\widehat{D} \to \widehat{D}_0 \neq 0$, $a' \to a'_\infty$, and the delay remainder satisfies $t^2\widehat{\varepsilon} \to 0$ in $L^2(\mathbb{R})$, the transverse metric components reduce asymptotically to
\begin{equation}
t^4 g^\perp_{\rm shift} \to \frac{C_*}{s_0^2}, \qquad
\frac{t^4 g^\perp_{\theta\theta}}{a_\infty'^2} \to \frac{C_*}{s_0^2},
\label{eq:prop1}
\end{equation}
where the universal constant is
\begin{equation}
C_* \equiv \left\| (1 - \Pi_{\widehat{D}_0}) \widehat{A}_0 \right\|_{L^2}^2 = \left\| (1 - \Pi_{\widehat{D}_0}) \widehat{Q} \right\|_{L^2}^2.
\label{eq:Cstar_def}
\end{equation}
Because $\widehat{A}_0$ is not parallel to $\widehat{D}_0$ and $a'_\infty \neq 0$, both limits are strictly positive, the ratio of transverse components satisfies $g^\perp_{\theta\theta}/g^\perp_{\rm shift} \to a_\infty'^2$, and the Fisher cosine between the corresponding residuals converges to unity.

\begin{table}[tbp]
\centering
\caption{Asymptotic predictions for FKPP ($\Delta x = 0.05$). The extrapolated limit $c_0$ of the fit $c_0 + c_1 t^{-1/2} + c_2 t^{-1} + c_3 t^{-3/2}$, for $t \ge 160$ and $t \ge 320$, is compared to $C_*/4 = 0.2780$. The coefficient $\partial_\theta\alpha_1$ is the $1/t$ coefficient of $\partial_\theta\sigma(t)$ (tangent equation, level $1/2$) for $t \ge 160$, and $t'_{0,\infty}$ is extrapolated linearly in $1/t$ from the tangent decomposition at $t = 160$--$1280$.}
\label{tab:graham}
\begin{ruledtabular}
\footnotesize
\setlength{\tabcolsep}{2.5pt}
\begin{tabular}{lccccc}
$\gamma_0$ & $t^4 g^\perp_{\rm shift}(1280)$ & $c_0$ & $\partial_\theta\alpha_1$ & $-\tfrac32 t'_{0,\infty}$ & $a'_\infty$ \\
\hline
$1.5$ & $0.2478$ & $0.2766$--$0.2773$ & $21.03$ & $20.89$ & $25.6$ \\
$2.0$ & $0.2497$ & $0.2764$--$0.2772$ & $2.371$ & $2.392$ & $2.34$ \\
$3.0$ & $0.2502$ & $0.2763$--$0.2772$ & $0.260$ & $0.263$ & $0.113$ \\
\end{tabular}
\end{ruledtabular}
\end{table}

\subsection{Input from the front asymptotics}
For step-like initial data ($u_0 = 1$ for $x < -L$, $u_0 = 0$ for $x > L$), Nolen, Roquejoffre and Ryzhik~\cite{Nolen:2017, Nolen:2019} and Graham~\cite{Graham:2017} proved that, with $\sigma$ given by \cref{eq:sigma},
\begin{equation}
u(x,t) = \phi(\xi) + \frac{\psi(\xi)}{t} + O\big(t^{-3/2+\gamma}\big), \qquad \xi = x - \sigma(t),
\label{eq:graham}
\end{equation}
locally uniformly in $\xi$ and for every $\gamma > 0$. Here $\psi$ does not depend on the initial data and solves
\begin{equation}
\psi'' + 2\psi' + (1-2\phi)\,\psi = \tfrac32\,\phi'.
\label{eq:psi}
\end{equation}
Graham's formal expansion to all orders~\cite[Sec.~5]{Graham:2017} gives the structure of the following terms. With $v = e^{\xi}u$, the FKPP equation becomes
\begin{equation}
v_t - v_{\xi\xi} + (\dot\sigma - 2)(v - v_\xi) + e^{-\xi} v^2 = 0 .
\end{equation}
At order $t^{-1}$, \cref{eq:psi} reads $\mathcal{L} V_1 = \tfrac32 (V_0 - V_0')$, with $V_0 = e^{\xi}\phi$, $V_1 = e^{\xi}\psi$ and $\mathcal{L} = -\partial_\xi^2 + 2e^{-\xi}V_0$. The coefficient $\alpha_1$ first enters $\dot\sigma - 2$ at order $t^{-2}$, through $d(\alpha_1/t)/dt = -\alpha_1/t^2$. The order-$t^{-2}$ equation therefore contains $\alpha_1$ only through the source $\alpha_1(V_0 - V_0')$. Its $\alpha_1$-dependent part is thus $\tfrac23 \alpha_1 V_1$, up to the kernel of $\mathcal{L}$, whose solutions that decay as $\xi \to -\infty$ are multiples of $e^{\xi}\phi'$, i.e., translations. The only other non-universal coefficient up to this order is $\alpha_0$. Hence, at fixed $\xi$,
\begin{equation}
\begin{aligned}
\partial_\theta u|_\xi &= \tfrac23\,\partial_\theta\alpha_1\, \frac{\psi}{t^2} + c(t)\,\phi' + o(t^{-2}), \\
\partial_t u|_\xi &= -\frac{\psi}{t^2} + o(t^{-2}).
\end{aligned}
\label{eq:timeshift}
\end{equation}
To this order, a change of initial condition is therefore a translation combined with the time shift $t_0' = -\tfrac23\partial_\theta\alpha_1$ of the universal relaxation $\psi/t$. Consistently, a temporal shift $t \mapsto t + T$ of a solution modifies the front position through the logarithmic term as
\begin{equation}
-\tfrac32\ln(t+T) = -\tfrac32\ln t - \frac{3T}{2t} + O(t^{-2}),
\end{equation}
changing the $1/t$ coefficient of $\sigma(t)$ by $\Delta\alpha_1 = -\tfrac32 T$. For an initial-condition variation producing an infinitesimal delay $dt_0 = t_0'\, d\theta$, this yields $\partial_\theta\alpha_1 = -\tfrac32 t_0'$, or equivalently $t_0' = -\tfrac23\partial_\theta\alpha_1$. Going to the laboratory frame and differentiating in $x$ gives the delay form, \cref{eq:delayform}, with
\begin{equation}
t_0' = -\tfrac23\,\partial_\theta\alpha_1, \qquad a' = \partial_\theta\sigma - \dot\sigma\, t_0' \to a'_\infty = \partial_\theta\alpha_0 + \tfrac43\,\partial_\theta\alpha_1,
\end{equation}
and $t^2\widehat{\varepsilon} \to 0$ in $L^2(\mathbb{R})$. Differentiating $p = -\partial_x u$ in the comoving frame yields $A = \partial_t p|_\xi \simeq \psi'/t^2 = -Q/t^2$, where $Q \equiv -\psi'$. Under the embedding, $t^2 \widehat{A} \to \widehat{A}_0 \equiv -Q/\sqrt{f} = -\widehat{Q}$, with $\widehat{Q} \equiv Q/\sqrt{f}$. Similarly, $\widehat{D} \to \widehat{D}_0 \equiv f'/\sqrt{f}$ with $f = -\phi'$, and $s_0 = 2$. \Cref{eq:prop1} then yields the asymptotic scaling law, \cref{eq:t4law}, with
\begin{equation}
C_* \equiv \left\| (1 - \Pi_{\widehat{D}_0}) \widehat{Q} \right\|_{L^2}^2 = \int_{-\infty}^{+\infty} \frac{\left[(1-\Pi_{f'})Q\right]^2}{f} \, d\xi,
\label{eq:Cstar_exact}
\end{equation}
where $\Pi_{f'}$ is the Fisher projection on $f'$. The embedded vector $\widehat{A}_0$ is not parallel to $\widehat{D}_0$: at the leading edge $\phi \sim B\xi e^{-\xi}$ while $\psi \sim -\tfrac{B}{4}\xi^3 e^{-\xi}$~\cite{Graham:2017}, so $Q/f'$ grows as $\xi^2$. The universal $t^{-3/2}$ term of the expansion gives relative corrections of order $t^{-1/2}$ to \cref{eq:t4law}.

\noindent\textit{Theoretical assumptions.} The scaling law, \cref{eq:t4law}, holds under three conditions: (i) it uses the formal $t^{-2}$ term of Graham's expansion~\cite{Graham:2017}, whereas published theorems bound the remainder only up to order $t^{-3/2+\gamma}$; (ii) differentiability of the expansion in $\theta$ and $t$, with convergence of the weighted Fisher integrals; (iii) extension from compact initial perturbations of a step to exponentially decaying data $u_0 = \min(1, e^{-\gamma_0 x})$ with $\gamma_0 > 1$, which select the same minimal speed. Under these assumptions, \cref{eq:t4law} is an exact consequence. Numerical verification is presented in \cref{fig:graham} and \cref{tab:graham}: the rescaled quantity $t^4 g^\perp_{\rm shift}$ converges to the same value for all initial conditions, and fitting its $O(t^{-1/2})$ approach in powers of $t^{-1/2}$ yields an extrapolated limit in sub-percent agreement with $C_*/4$. The transverse residual converges to the universal profile $(1-\Pi_{f'})Q$ (Fisher cosine $0.99975$ at $t = 320$ and $0.99995$ at $t = 1280$ for all three $\gamma_0$), and the delay relation $\partial_\theta\alpha_1 = -\tfrac32 t_0'$ is confirmed to within $1.2\%$. The result is therefore a conditional asymptotic derivation, verified numerically, rather than a general proof of the $t^{-4}$ law.

\bibliography{references}

@article{Munier:2003gx,
    author = "Munier, S. and Peschanski, R.",
    title = "{Geometric scaling as traveling waves}",
    eprint = "hep-ph/0309177",
    archivePrefix = "arXiv",
    doi = "10.1103/PhysRevLett.91.232001",
    journal = "Phys. Rev. Lett.",
    volume = "91",
    pages = "232001",
    year = "2003"
}

@article{Munier:2003sj,
    author = "Munier, S. and Peschanski, R.",
    title = "{Traveling wave fronts and the transition to saturation}",
    eprint = "hep-ph/0310357",
    archivePrefix = "arXiv",
    doi = "10.1103/PhysRevD.69.034008",
    journal = "Phys. Rev. D",
    volume = "69",
    pages = "034008",
    year = "2004"
}

@article{Balitsky:1995ub,
    author = "Balitsky, I.",
    title = "{Operator expansion for high-energy scattering}",
    eprint = "hep-ph/9509348",
    archivePrefix = "arXiv",
    doi = "10.1016/0550-3213(95)00638-9",
    journal = "Nucl. Phys. B",
    volume = "463",
    pages = "99--160",
    year = "1996"
}

@article{Kovchegov:1999yj,
    author = "Kovchegov, Yuri V.",
    title = "{Small x F(2) structure function of a nucleus including multiple pomeron exchanges}",
    eprint = "hep-ph/9901281",
    archivePrefix = "arXiv",
    doi = "10.1103/PhysRevD.60.034008",
    journal = "Phys. Rev. D",
    volume = "60",
    pages = "034008",
    year = "1999"
}

@article{Kovchegov:2000hz,
    author = "Kovchegov, Yuri V.",
    title = "{Unitarization of the BFKL pomeron on a nucleus}",
    eprint = "hep-ph/9905214",
    archivePrefix = "arXiv",
    doi = "10.1103/PhysRevD.61.074018",
    journal = "Phys. Rev. D",
    volume = "61",
    pages = "074018",
    year = "2000"
}

@article{Fisher:1937,
    author = "Fisher, R. A.",
    title = "{The wave of advance of advantageous genes}",
    journal = "Ann. Eugenics",
    volume = "7",
    number = "4",
    pages = "355--369",
    year = "1937",
    doi = "10.1111/j.1469-1809.1937.tb02153.x"
}

@article{Kolmogorov:1937,
    author = "Kolmogorov, A. and Petrovsky, I. and Piscounov, N.",
    title = "{{\'E}tude de l'{\'e}quation de la diffusion avec croissance de la quantit{\'e} de mati{\`e}re et son application {\`a} un probl{\`e}me biologique}",
    journal = "Bull. Univ. Moscou, S{\'e}r. Int. A",
    volume = "1",
    number = "6",
    pages = "1--26",
    year = "1937"
}

@article{Stasto:2000er,
    author = "Sta{\'s}to, A. M. and Golec-Biernat, K. and Kwieci{\'n}ski, J.",
    title = "{Geometric scaling for the total $\gamma^* p$ cross-section in the low $x$ region}",
    eprint = "hep-ph/0007192",
    archivePrefix = "arXiv",
    doi = "10.1103/PhysRevLett.86.596",
    journal = "Phys. Rev. Lett.",
    volume = "86",
    pages = "596--599",
    year = "2001"
}

@article{Bramson:1983,
    author = "Bramson, Maury",
    title = "{Convergence of solutions of the Kolmogorov equation to travelling waves}",
    journal = "Mem. Amer. Math. Soc.",
    volume = "44",
    number = "285",
    pages = "1--190",
    year = "1983",
    doi = "10.1090/memo/0285"
}

@article{vanSaarloos:2000,
    author = "Ebert, U. and van Saarloos, W.",
    title = "{Front propagation into unstable states: Universal algebraic convergence towards uniformly translating pulled fronts}",
    doi = "10.1016/S0167-2789(00)00068-3",
    journal = "Physica D",
    volume = "146",
    number = "1-4",
    pages = "1--99",
    year = "2000"
}

@article{Berestycki:2017,
    author = "Berestycki, J. and Brunet, {\'E}. and Derrida, B.",
    title = "{Exact solution and precise asymptotics of a Fisher--KPP type front}",
    eprint = "1705.08416",
    archivePrefix = "arXiv",
    primaryClass = "math.PR",
    doi = "10.1088/1751-8121/aa899f",
    journal = "J. Phys. A: Math. Theor.",
    volume = "51",
    number = "3",
    pages = "035204",
    year = "2018"
}

@article{Graham:2017,
    author = "Graham, Cole",
    title = "{Precise asymptotics for Fisher-KPP fronts}",
    eprint = "1712.02472",
    archivePrefix = "arXiv",
    primaryClass = "math.AP",
    doi = "10.1088/1361-6544/aaffe8",
    journal = "Nonlinearity",
    volume = "32",
    number = "6",
    pages = "1967--1998",
    year = "2019"
}

@article{Peschanski:2012cw,
    author = "Peschanski, Robi",
    title = "{Dynamical entropy of dense QCD states}",
    eprint = "1211.6911",
    archivePrefix = "arXiv",
    primaryClass = "hep-ph",
    doi = "10.1103/PhysRevD.87.034042",
    journal = "Phys. Rev. D",
    volume = "87",
    number = "3",
    pages = "034042",
    year = "2013"
}

@article{Ramos:2022,
    author = "Ramos, G. S. and Machado, M. V. T.",
    title = "{Investigating the QCD dynamical entropy in high-energy hadronic collisions}",
    eprint = "2203.10986",
    archivePrefix = "arXiv",
    primaryClass = "hep-ph",
    doi = "10.1103/PhysRevD.105.094009",
    journal = "Phys. Rev. D",
    volume = "105",
    number = "9",
    pages = "094009",
    year = "2022"
}

@article{Ramos:2025,
    author = "Ramos, G. S. and Moriggi, L. S. and Machado, M. V. T.",
    title = "{Investigating QCD dynamical entropy in high-energy nuclear collisions}",
    eprint = "2507.09349",
    archivePrefix = "arXiv",
    primaryClass = "hep-ph",
    doi = "10.1016/j.physletb.2025.139737",
    journal = "Phys. Lett. B",
    volume = "868",
    pages = "139737",
    year = "2025"
}

@unpublished{Kou:2026dglap,
    author = "Kou, Wei and Chen, Xurong",
    title = "{Information Geometry of DGLAP Evolution: From Geodesic Tests to Emergent Small-Circle Geometry}",
    eprint = "2610.27886",
    archivePrefix = "arXiv",
    primaryClass = "hep-ph",
    year = "2026"
}

@unpublished{Bondarenko:2026dglap,
    author = "Bondarenko, S. and Prygarin, A.",
    title = "{Information geometry of DGLAP evolution}",
    eprint = "2610.28115",
    archivePrefix = "arXiv",
    primaryClass = "hep-ph",
    year = "2026"
}

@article{Ito:2020,
    author = "Ito, Sosuke and Dechant, Andreas",
    title = "{Stochastic time evolution, information geometry, and the Cram{\'e}r-Rao bound}",
    eprint = "1810.06832",
    archivePrefix = "arXiv",
    primaryClass = "cond-mat.stat-mech",
    doi = "10.1103/PhysRevX.10.021056",
    journal = "Phys. Rev. X",
    volume = "10",
    number = "2",
    pages = "021056",
    year = "2020"
}

@article{GolecBiernat:1998js,
    author = "Golec-Biernat, K. and W{\"u}sthoff, M.",
    title = "{Saturation effects in deep inelastic scattering at low $Q^2$ and its implications on diffraction}",
    eprint = "hep-ph/9807513",
    archivePrefix = "arXiv",
    doi = "10.1103/PhysRevD.59.014017",
    journal = "Phys. Rev. D",
    volume = "59",
    pages = "014017",
    year = "1998"
}

@article{McLerran:1993ni,
    author = "McLerran, Larry D. and Venugopalan, Raju",
    title = "{Computing quark and gluon distribution functions for very large nuclei}",
    eprint = "hep-ph/9309289",
    archivePrefix = "arXiv",
    doi = "10.1103/PhysRevD.49.2233",
    journal = "Phys. Rev. D",
    volume = "49",
    pages = "2233--2241",
    year = "1994"
}

@article{McLerran:1994ka,
    author = "McLerran, Larry D. and Venugopalan, Raju",
    title = "{Gluon distribution functions for very large nuclei at small transverse momentum}",
    eprint = "hep-ph/9311205",
    archivePrefix = "arXiv",
    doi = "10.1103/PhysRevD.49.3352",
    journal = "Phys. Rev. D",
    volume = "49",
    pages = "3352--3355",
    year = "1994"
}

@article{Kuraev:1977fs,
    author = "Kuraev, E. A. and Lipatov, L. N. and Fadin, V. S.",
    title = "{The Pomeranchuk singularity in non-Abelian gauge theories}",
    journal = "Sov. Phys. JETP",
    volume = "45",
    pages = "199--204",
    year = "1977"
}

@article{Balitsky:1978,
    author = "Balitsky, I. I. and Lipatov, L. N.",
    title = "{The Pomeranchuk singularity in quantum chromodynamics}",
    journal = "Sov. J. Nucl. Phys.",
    volume = "28",
    pages = "822--829",
    year = "1978"
}

@article{Mueller:2002zm,
    author = "Mueller, A. H. and Triantafyllopoulos, D. N.",
    title = "{The energy dependence of the saturation momentum}",
    eprint = "hep-ph/0205167",
    archivePrefix = "arXiv",
    doi = "10.1016/S0550-3213(02)00581-3",
    journal = "Nucl. Phys. B",
    volume = "640",
    pages = "331--350",
    year = "2002"
}

@article{Levin:2000,
    author = "Levin, E. and Tuchin, K.",
    title = "{Solution to the evolution equation for high parton density QCD}",
    eprint = "hep-ph/9908317",
    archivePrefix = "arXiv",
    doi = "10.1016/S0550-3213(99)00825-1",
    journal = "Nucl. Phys. B",
    volume = "573",
    pages = "833--852",
    year = "2000"
}

@article{Albacete:2010sy,
    author = "Albacete, Javier L. and Armesto, Nestor and Milhano, Jose Guilherme and Quiroga-Arias, Paloma and Salgado, Carlos A.",
    title = "{AAMQS: A non-linear QCD analysis of new HERA data at small x including heavy quarks}",
    eprint = "1012.4408",
    archivePrefix = "arXiv",
    primaryClass = "hep-ph",
    doi = "10.1140/epjc/s10052-011-1705-3",
    journal = "Eur. Phys. J. C",
    volume = "71",
    pages = "1705",
    year = "2011"
}

@article{Lappi:2013zma,
    author = "Lappi, T. and M{\"a}ntysaari, H.",
    title = "{Single inclusive particle production at high energy from HERA data to proton-nucleus collisions}",
    eprint = "1309.6963",
    archivePrefix = "arXiv",
    primaryClass = "hep-ph",
    doi = "10.1103/PhysRevD.88.114020",
    journal = "Phys. Rev. D",
    volume = "88",
    number = "11",
    pages = "114020",
    year = "2013"
}

@article{Nolen:2017,
    author = "Nolen, J. and Roquejoffre, J.-M. and Ryzhik, L.",
    title = "{Convergence to a single wave in the Fisher-KPP equation}",
    eprint = "1604.02994",
    archivePrefix = "arXiv",
    primaryClass = "math.AP",
    doi = "10.1007/s11401-017-1087-4",
    journal = "Chin. Ann. Math. Ser. B",
    volume = "38",
    number = "2",
    pages = "629--646",
    year = "2017"
}

@article{Nolen:2019,
    author = "Nolen, J. and Roquejoffre, J.-M. and Ryzhik, L.",
    title = "{Refined long-time asymptotics for Fisher-KPP fronts}",
    eprint = "1607.08802",
    archivePrefix = "arXiv",
    primaryClass = "math.AP",
    doi = "10.1142/S0219199718500724",
    journal = "Commun. Contemp. Math.",
    volume = "21",
    number = "7",
    pages = "1850072",
    year = "2019"
}

@book{Amari:2000,
    author = "Amari, Shun-ichi and Nagaoka, Hiroshi",
    title = "{Methods of Information Geometry}",
    series = "Translations of Mathematical Monographs",
    volume = "191",
    publisher = "American Mathematical Society",
    address = "Providence, RI",
    doi = "10.1090/mmono/191",
    year = "2000"
}

\end{document}